\documentstyle[prb,aps,epsfig,amssymb]{revtex}

\begin{document}

\title{Reply to ``Comment on `Origin of combination frequencies in quantum magnetic oscillations of two-dimensional multiband metals' '' by A.S. Alexandrov and A.M. Bratkovsky [cond-mat/0207173]}
\author{T. Champel}
\address{
Commissariat \`{a} l'Energie Atomique,  DSM/DRFMC/SPSMS\\
17 rue des Martyrs, 38054 Grenoble Cedex 9, France}
\date{\today}

\maketitle

\begin{abstract}
In their comment on the paper (Phys. Rev. B 65, 153403 (2002);
cond-mat/0110154), Alexandrov and Bratkovsky (cond-mat/0207173)
argue that they correctly took into account the chemical potential
oscillations in their analytical theory of combination frequencies
in multiband low-dimensional metals by expanding the free energy
in powers of the chemical potential oscillations. In this reply,
we show that this claim contradicts their original paper (Phys.
Rev. B 63, 033105 (2001)). We demonstrate that the condition given
for the expansion is mathematically incorrect. The correct
condition allows to understand the limits of validity of the
analytical theory.

\end{abstract}


It is well-known that, due to the quantization of the electronic energy spectrum of metals into discrete Landau levels in the presence of a magnetic field, the chemical potential is expected to oscillate with the magnetic field when the number of electrons is kept constant. In two-dimensional (2D) multiband metals, it has been  predicted initially in the framework of numerical studies~\cite{Ale1996,Nak1999} that such chemical potential oscillations are responsible for the appearance of additional oscillations of the magnetization
whose frequencies are combinations of the independent band frequencies $f_{\alpha}$.
The number of electrons is fixed experimentally, independently of the dimensionality of the energy spectrum.
For this reason, combination frequencies are in principle possible as well
in 3D metals as in 2D multiband metals.
However,
chemical potential oscillations effects are not observed in multiband 3D metals. So, the mechanism of appearance of combination frequencies has to  clearly point out the difference between low-dimensional metals and 3D metals.

Alexandrov and Bratkovsky are the first authors to propose an analytical derivation for these combination frequencies~\cite{Ale2001}.
However, they did not mention clearly the difference between low-dimensional metals and 3D metals in the framework of their theory. More precisely, an important point noticed for one-band 2D metals was missing in their derivation \cite{Ale2001}: the analytical treatment of the chemical potential oscillations effects leads to a system of  nonlinear equations~\cite{Cha2001,Gri2001}. In 3D metals, the resolution of these equations is trivial because the oscillating part $\tilde{\mu}$ of the chemical potential is of the order of $\hbar\omega_{c} \sqrt{\hbar \omega_{c}/\varepsilon_{F}}$ (where $\omega_{c}$ is the cyclotron pulsation and $\varepsilon_{F}$ the Fermi energy). As a result, the
 magnetization oscillations are not sensitive to the chemical potential oscillations effects: the expression for the magnetization is (with high accuracy) the same for a fixed chemical potential $\mu$ as for a fixed number of electrons $N$.

In 2D metals $\tilde{\mu} \sim \hbar \omega_{c}$, that leads to
 the observable difference between low-dimensional metals and 3D metals.
This fact has not been noticed in the article~\cite{Ale2001}
and has motivated the analytical work~\cite{Cha2002} and its principal statement that the chemical potential oscillations appearing in the arguments of the Fourier components (of the grand canonical potential or of the magnetization oscillations) were not taken into account by Alexandrov and Bratkovsky.

In their comment~\cite{Ale2003}, these authors have addressed principally two criticisms to the paper \cite{Cha2002}. First, (i) the chemical potential oscillations  were correctly taken into account in their previous work \cite{Ale2001} :
they argue that they made in fact an expansion (as in the commented paper \cite{Cha2002}), but did not judge to mention explicitly this technical point~\cite{Ale2003}.
Secondly, (ii) the analytical formula for the combination frequencies amplitude they derived is accurate even at zero temperature in a clean two-band 2D metal.
We will show in this reply that the argument (i) is in contradiction with the original paper \cite{Ale2001} and  is thus not receivable.
The point (ii)  is far from being correct.

The main point of the Ref. \cite{Ale2001} is to express the oscillating part $\tilde{F}$ of the free energy as
\begin{equation}
\tilde{F}
=\tilde{\Omega}(\mu)
-\frac{1}{2 \rho}
 \left[\left( \frac{\partial \tilde{\Omega}}{\partial \mu} \right)_{H}(\mu)\right]^{2}
\end{equation}
where $\rho$ is the total density of states of a two-dimensional multiband metal, and $\tilde{\Omega}(\mu)$ is the oscillating part of the grand canonical potential, which is an explicit function of the chemical potential $\mu$.
At a constant number of electrons $N$, the chemical potential $\mu$ oscillates with the magnetic field and can be written as the sum of a constant part $\mu_{0}$ (independent of the magnetic field) and an oscillating part $\tilde{\mu}$ given by

\begin{equation}
\tilde{\mu}=\frac{1}{\rho}
\left(
\frac{
\partial
\tilde{\Omega}
}
{\partial \mu} \right)_{H}(\mu)=\mu-\mu_{0}
\end{equation}
As stressed in the Ref. \cite{Cha2002}, this equation (2) is a self-consistent nonlinear equation to solve in order to determine the dependence of the chemical potential $\mu$ on the magnetic field $H$.

The grand canonical potential $\tilde{\Omega}(\mu)$ depends on the chemical potential $\mu$ through the expression

\begin{equation}
\tilde{\Omega}(\mu)=\sum_{\alpha} \sum_{r=1}^{\infty} (-1)^{l}A_{\alpha}^{r} \cos\left(2\pi r \frac{\mu_{\alpha}}{\hbar \omega_{c\alpha}}\right)
\end{equation}
where $$
2\pi \frac{\mu_{\alpha}}{\hbar \omega_{c \alpha}}=2 \pi \frac{\mu-\Delta_{\alpha}}{\hbar \omega_{c \alpha}}=\frac{f_{\alpha}}{H}$$
are the arguments entering in the Fourier components.
Here $\omega_{c\alpha}$ is the cyclotron pulsation with the effective mass $m_{\alpha}$, $\Delta_{\alpha}$ is the $\alpha$ band-edge, and
$$
A_{\alpha}^{r}=\frac{m_{\alpha}\omega_{c\alpha}}{2\pi^{3}}R_{T}(r)R_{D}(r)
$$
 is the amplitude for the harmonic $r$ in the  band $\alpha$ with
$$R_{T}(r)=\frac{\lambda_{r}}{\sinh \lambda_{r}} , \hspace{1cm} R_{D}(r)=\exp\left(-2\pi r \frac{\Gamma_{0}}{\hbar \omega_{c\alpha}}\right),$$
$T$ the temperature, $\lambda_{r}=2\pi^{2}r k_{B}T/\hbar \omega_{c\alpha}$, and $\Gamma_{0}$ the relaxation rate at $H=0$.
Since for a constant number of electrons the chemical potential oscillates with  the magnetic field, so does the quantity $f_{\alpha}$. For this reason, the explicit expression (3) can not be seen as a Fourier series ($f_{\alpha}$ is not a frequency, which by definition has to be independent of the magnetic field).

This difficulty can be overcomed by expanding the free energy $\tilde{F}(\mu)$  (expression (1)) in powers of the oscillating part $\tilde{\mu}$ of the chemical potential.
By developing separately the first term and the last term of the right-hand side of the equation (1), we straightforwardly obtain keeping terms up to the order of $\rho \tilde{\mu}^{2}$
\begin{equation}
\tilde{\Omega}(\mu) \approx \tilde{\Omega}(\mu_{0})+\left(\mu -\mu_{0}\right) \left( \frac{\partial \tilde{\Omega}}{\partial \mu} \right)_{H}(\mu_{0})
=\tilde{\Omega}(\mu_{0})+\frac{1}{\rho}\left[\left( \frac{\partial \tilde{\Omega}}{\partial \mu} \right)_{H}(\mu_{0}) \right]^{2}
\end{equation}
and
\begin{equation}
-\frac{1}{2\rho}\left[\left( \frac{\partial \tilde{\Omega}}{\partial \mu} \right)_{H}(\mu)\right]^{2}\approx
-\frac{1}{2\rho}\left[
 \left( \frac{\partial \tilde{\Omega}}{\partial \mu} \right)_{H}(\mu_{0}) \right]^{2}.
\end{equation}
Therefore the free energy becomes
\begin{equation}
\tilde{F}
=\tilde{\Omega}(\mu_{0})
+\frac{1}{2 \rho}
 \left[\left( \frac{\partial \tilde{\Omega}}{\partial \mu} \right)_{H}(\mu_{0}) \right]^{2}.
\end{equation}
Obviously after equations (4) and (5), Fourier harmonics with combination harmonics are produced by both terms of the right-hand side of the expression (1), the contribution of the first term being reduced to one half due to a partial cancellation with the second term.

Alexandrov and Bratkovsky argue~\cite{Ale2003} that the free energy $\tilde{F}$ is expanded in powers of $\tilde{\mu} \ll \mu_{0}$ in their original work \cite{Ale2001}; in fact,
 the authors asserted just after giving the expression (1) (their Eq. (12)) in the Ref.~\cite{Ale2001} : ``{\em It is the last term, which yields combination Fourier harmonics with the combination frequencies $f=rf_{\alpha}\pm r'f_{\alpha'}$}''.
Thus, this sentence directly contradicts the fact that the free energy has been  expanded in the paper~\cite{Ale2001}: the chemical potential oscillations were explicitly not taken into account in the trigonometric arguments.

It is worth noting that a similar claim
 has been addressed independently by Kishigi and Hasegawa~\cite{Kis2002}, who wrote concerning the authors of Ref. \cite{Ale2001}:``{\em their result for the free energy [...] is formally correct but they did not take account of the magnetic-field dependence of the $f_{\alpha}$, which cannot be neglected in two-dimensional systems. As a result their analysis of the de Haas-van Alphen oscillation for the fixed $N$ system (canonical ensemble) is insufficient and their conclusions on Fourier-transform intensities are incorrect}''.

The similitude of the formulae obtained in Ref.~\cite{Cha2002} and Ref.~\cite{Ale2001} has a simple explanation. We note by comparing the expressions (1) and (6) that the squared term changes sign after the expansion.
It implies that if we ignore by hand the oscillations of the chemical potential  in the quantity $f_{\alpha}$, the amplitude obtained for the combination harmonics is fortuitously the same (a sign apart) after and before the development.

The second point (ii) of discordance is in fact related to the validity of
 the analytical expansion of $\tilde{F}$ (or of $\tilde{\Omega}$, or of the magnetization oscillations $\tilde{M}$) in powers of the oscillating part $\tilde{\mu}$
 of the chemical potential $\mu$.
The condition of validity for it, $|\tilde{\mu}| \ll \mu_{0}$ (which is fulfilled in the whole regime of magnetic quantum oscillations) given in~\cite{Ale2003}, is actually mathematically incorrect. The correct condition for performing the developments (4-6) is revealed when considering the explicit form (3) for $\tilde{\Omega}(\mu)$. The basic point is that the oscillating part $\tilde{\mu}$ enters in the arguments of trigonometric functions for which
the approximation
$$\cos (x_{0}+\tilde{x}) \approx \cos(x_{0})$$
is valid under the condition $|\tilde{x}| \ll 1$ (and not $|\tilde{x}| \ll x_{0}$, take e.g. $\pi/2 \ll x_{0}$).
The expansion (6)  is therefore valid provided that
\begin{equation}
2 \pi r \left|\frac{\tilde{\mu}}{\hbar \omega_{c\alpha}}\right| \ll 1
\end{equation}
for all significant harmonics $r$. Obviously, this latter condition (7) is much more restrictive than the condition given by the authors of the Ref. \cite{Ale2003} and depends on the degree $r$ of the harmonic (it is stronger for higher harmonics).  It holds simultaneously for the expansion of all quantities which are functions of $\mu/\omega_{c \alpha}$ through trigonometric arguments such as $\tilde{\Omega}$, the magnetization oscillations $\tilde{M}$ or even $\tilde{\mu}$ itself (see equation (2)).

It has already been noticed in Ref. \cite{Cha2002} that $\tilde{\mu}$ is naturally reduced by the presence of multiple bands independently of the temperature or the impurity reduction factors. However, at very low temperatures and in clean 2D two-band metals (the most unfavorable case), many harmonics $r$ are significant, and the condition (7) is not fulfilled for all terms
because in this case $|\tilde{\mu}|$  is of the order of a few tenths of $\hbar\omega_{c \alpha}$. Then, higher powers of $\tilde{\mu}$ have to be considered and the analytical treatment of the chemical potential oscillations effects is not obvious. In this regime of so-called strong chemical potential oscillations the validity of the analysis of the oscillations in terms of Fourier series is questioned at small but finite temperatures (this is not a property established a priori; the use of the Fourier analysis has thus to be justified).

The importance of higher powers of $\tilde{\mu}$ has been demonstrated analytically  and numerically in the case of one-band 2D metals~\cite{Cha2001,Gri2001}. In these papers~\cite{Cha2001,Gri2001} the full nonlinear equation (2) is considered and solved at zero temperature. As a result, the drops of the magnetization oscillations occur at integer values of the ratio $\mu_{0}/\hbar \omega_{c}$, while they occur at half-integer values when neglecting $\tilde{\mu}$ in the trigonometric arguments. Furthermore, harmonic amplitudes are found to differ strongly in the two situations, especially for high harmonics~\cite{Gri2001}.

In the two-band 2D metals, the combination frequencies appear in the first order expansion in powers of $\tilde{\mu}$. Alexandrov and Bratkovsky argue~\cite{Ale2003} with the help of an analytical and a numerical estimates that higher powers of $\tilde{\mu}$ can be neglected to describe the magnetization oscillations with a relatively good accuracy even at zero temperature and in clean samples.
In their analytical estimate, they only
keep the first harmonic in the self-consistent equation (2) for the chemical potential oscillations, which is valid in fact at zero temperature only at a small Dingle factor $R_{D}(1)$.
In this regime, the oscillations of $\tilde{\mu}$ are small and particularly smooth: for this reason the linear approximation is naturally expected to be quite good since the condition (7) is then fulfilled.
On the contrary, in their numerical study they  consider all the harmonics in Eq. (2), which is relevant in the regime of strong oscillations of
$\tilde{\mu}$.
We can however cast some doubts on their result shown in the Fig. 1 of Ref. \cite{Ale2003}: surprisingly, at small Dingle factors (i.e. when the expansion is possible), the accuracy of the linear approximation is increasing  when the Dingle factor $R_{D}(1)$ decreases; this is completely opposite to that can be expected.

In conclusion, we have demonstrated that the chemical potential oscillations were not correctly taken into account in the Ref. \cite{Ale2001}. Careful numerical studies are still needed to analyze the regime of strong chemical potential oscillations where the expansion of the quantities in powers of $\tilde{\mu}$ is not convergent.

I thank V.P. Mineev for his advices and careful reading of the manuscript.

\end{document}